\documentclass{sig-alternate}

\usepackage{amsfonts, amssymb, amsmath}
\usepackage{graphicx}
\usepackage{epsfig}
\usepackage{subfigure}
\usepackage{xspace}
\usepackage{mathptmx}
\usepackage[ruled]{algorithm2e}
\usepackage{ctable,multirow}
\SetKwInOut{In}{input}\SetKwInOut{Out}{output}
\usepackage{url}

\newtheorem{lemma}{Lemma}
\newtheorem{definition}{Definition}

\newcommand{\Rea}{\mathbb{R}}
\newcommand{\eat}[1]{}
\newcommand{\anv}{attribute $\langle$name, value$\rangle$ }

\newcommand{\eg}{\emph{e.g., }}

\newcommand{\av}[1]{(\textrm{#1})}
\newcommand{\query}[1]{[\texttt{#1}]}
\newcommand{\Prob}{\ensuremath\mathbb{P}}

\conferenceinfo{CIKM'12,} {Oct 29 - Nov 2, 2012, Maui, Hawaii.}
\CopyrightYear{2012}
\title{Structured Query Reformulations in Commerce Search}
\numberofauthors{1}
\author{}
\author{\alignauthor
Sreenivas Gollapudi \qquad
Samuel Ieong \qquad
Anitha Kannan \\
\affaddr{Microsoft Research Silicon Valley}\\
\email{\{sreenig,saieong,ankannan\}@microsoft.com}
}

\begin{document}

\maketitle

\begin{abstract}
Recent work in commerce search has shown that understanding the semantics in user queries enables more effective query analysis and retrieval of relevant products. However, due to lack of sufficient domain knowledge, user queries often include terms that cannot be mapped directly to any product attribute.  For example, a user looking for {\tt designer handbags} might start with such a query because she is not familiar with the manufacturers, the price ranges, and/or the material that gives a handbag designer appeal. Current commerce search engines treat terms such as {\tt designer} as keywords and attempt to match them to contents such as product reviews and product descriptions, often resulting in poor user experience.

In this study, we propose to address this problem by reformulating queries involving terms such as {\tt designer}, which we call \emph{modifiers}, to queries that specify precise product attributes.  We learn to rewrite the modifiers to attribute values by analyzing user behavior and leveraging structured data sources such as the product catalog that serves the queries. We first produce a probabilistic mapping between the modifiers and attribute values based on user behavioral data. These initial associations are then used to retrieve products from the catalog, over which we infer sets of attribute values that best describe the semantics of the modifiers.  We evaluate the effectiveness of our approach based on a comprehensive Mechanical Turk study.  We find that users agree with the attribute values selected by our approach in about 95\% of the cases and they prefer the results surfaced for our reformulated queries to ones for the original queries in 87\% of the time.
\end{abstract}


\section{Introduction} \label{sec:intro}
There has been tremendous growth in the amount of commerce conducted over the web in the past decade.  In a recent survey, ComScore  reported a record-breaking \$44.3 billion e-commerce retail spending in the U.S. in the first quarter of 2012, up 17\% from the previous year~\cite{comscore}.  Nearly 70\% of all Internet users have made at least one online purchase during this time.

Whether through dedicated e-commerce sites such as Amazon or search engine verticals such as Google or Bing shopping, most online transactions begins with \emph{search}. However, there are important differences between web search and commerce search.  While web search is typically performed over unstructured data such as contents of webpages, commerce search is typically performed over structured data in the form of a product catalog.  The catalog provides rich semantics that can be associated with both queries and products, and can lead to more effective query analysis and ranking~\cite{GINP11,SPT10}.  For example, techniques exist to annotate keywords queries with type semantics, such as annotating \query{nikon digital cameras} as \query{brand:nikon category:digital cameras}~\cite{SPT10}.  These queries can then be used to search structured data sources and enable search engines to find more relevant products.

However, not all commerce queries can be annotated with such clean semantics.  Due to possible lack of domain knowledge, users often express their information needs with query terms that cannot be directly annotated using the structured data in the product catalog. For example, consider the query \query{designer handbags}.  A user may issue this query to discover aspects such as brands and materials that constitute ``designer'' appeal to handbags, and expect the search engine to retrieve products that capture these nuances.  However, there is no explicit type semantics that can be associated with terms such as \query{designer} based on the catalog, leaving such terms as \emph{free tokens} (as opposed to \emph{typed tokens} such as \query{brand:nikon}) to be handled by the retrieval system.

Current commerce search engine approaches this challenge by matching the free tokens as keywords.  For this to work, the search engine must decide on the sources of information over which the matching is done.  Typical sources include product descriptions and user reviews. There are several drawbacks to this solution.  First, these sources could be noisy---a seller may have incentives to label all of the handbags with positive terms such as ``designer'' and ``stylish'' to boost sales.  Second, the information could be dated---a handbag that is considered designer a year ago may become blas\'{e} today.  Finally, relying on textual matches could adversely affect recall in cases where the free token is rare.

\begin{figure}[t]
\centering
\includegraphics[width=\columnwidth]{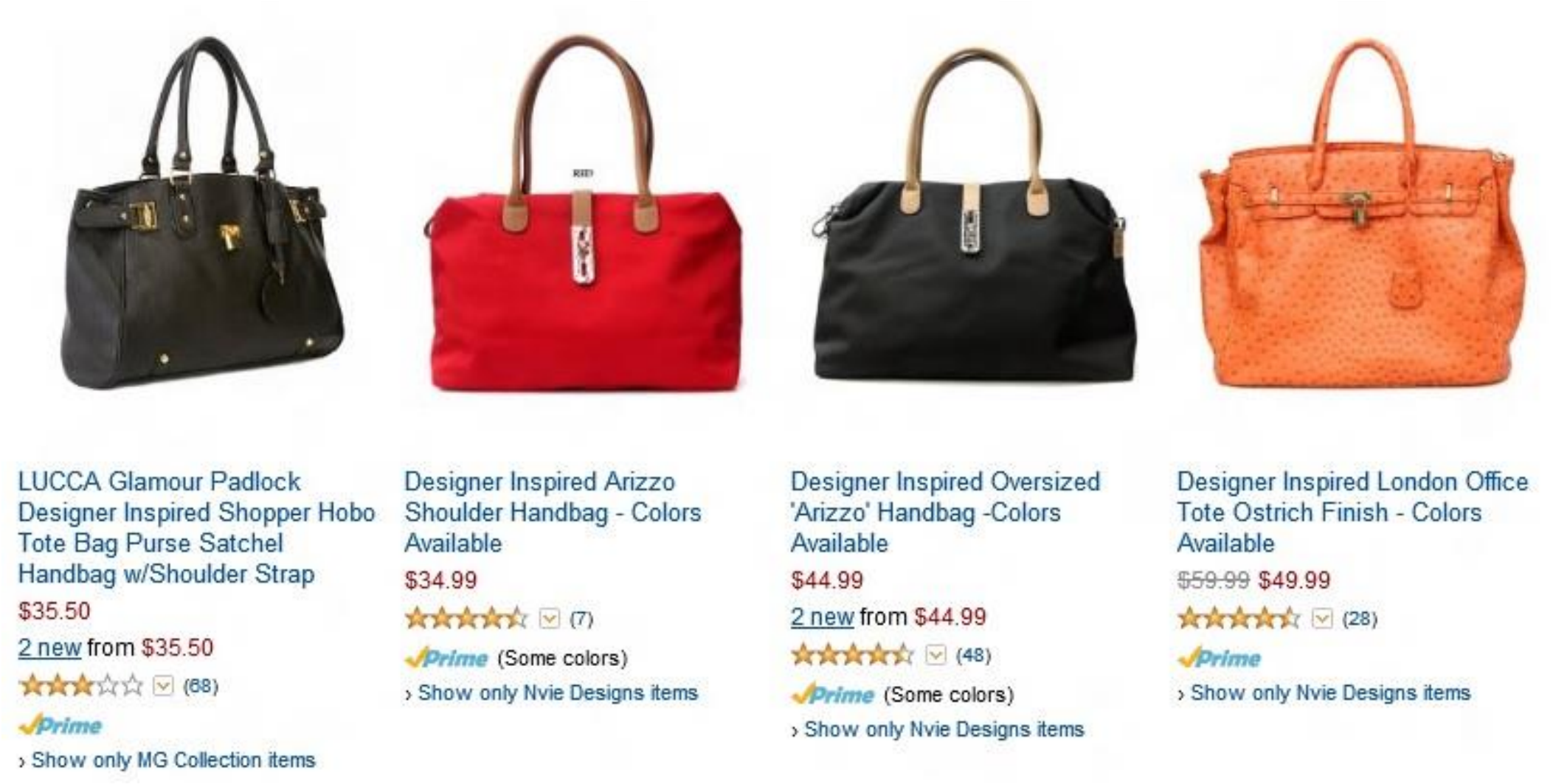}
\caption{Results for the query \query{designer handbags} on Amazon}
\label{fig:dh}
\end{figure}

Figure~\ref{fig:dh} shows an illustrative example that demonstrates the limitation of matching free tokens as keywords. In this figure, we show the results for the query \query{designer handbags} on Amazon, with search restricted to category {\tt Women's Handbags \& Purses} (results without restriction appear worse).  In our opinion, the retrieved products are poor reflections of characteristics of ``designer'' handbags.  These products are likely retrieved due to the term ``designer'' in their titles.  We believe better results can be retrieved through query reformulation, for example, by reformulating the query \query{designer handbags} to \query{gucci leather handbags}.

There are many ways in which free tokens can change the intents of the queries, and thus the set of results to be retrieved.  Our work focuses on free tokens that change the set of products to be retrieved as opposed to ones that change the nature of the products to be retrieved.  We call the former \emph{modifiers}.  For example, in the category of televisions, the free token \texttt{portable} indicates that the users are likely looking for televisions that are light in weight and small in size, whereas the free token \texttt{manual} indicate that the users are looking for operation instructions for certain televisions and not televisions themselves.

In this study, we propose a principled approach to reformulate queries containing modifiers to ones that specify precise attributes.  Our intuition is that users who issue queries such as \query{designer handbags} will end up spending more time \emph{browsing} \query{designer handbags} in their search session, and can thus inform us the attributes that give handbags ``designer'' appeal.
We first learn to identify the set of modifiers among all free tokens.  We then learn the association of different attribute values with modifiers by analyzing the user browsing behavior in search sessions.  These associations are then used to retrieve products from the product catalog, over which we infer sets of attribute values that best describe the modifiers.  We conduct a comprehensive study to evaluate our approach, and find that in 95\% of the cases users agree with our selected attribute values for the modifiers and in 87\% of the cases prefer the results based on our reformulated queries to the original ones.

The remainder of the paper is organized as follows.  We first review past work on commerce search and query reformulation in Section~\ref{sec:related}.  We then explain our approach of combining user browse signals with structured data from product catalogs to produce candidate reformulations in Section~\ref{sec:model}.  We present a comprehensive user study conducted over the Mechnical Turk platform in Section~\ref{sec:experiments}.  We conclude with our main findings and suggest future research directions in Section~\ref{sec:conclusion}.
%

\section{Related Work} \label{sec:related}
Our work is motivated by the problem of answering commerce queries against a product catalog used in a commercial shopping search engine such as those used in Amazon and Bing. As mentioned in the Introduction, a commerce query can be thought of being composed of semantic tokens that can be annotated with corresponding attributes, which we call \emph{typed tokens}, and \emph{free tokens}. As an example, for the query \query{prada designer handbags}, the typed tokens are \query{brand:prada category:handbags} while the free token is \query{designer}.  Much of the work in literature has focused on inferring the attributes to be associated with typed tokens. In \cite{LWA09}, a conditional random field is trained to infer attributes corresponding to query tokens,  while  in \cite{SPT10}, a probabilistic generative model is trained to infer the most likely complete annotation for the query. In both these methods, tokens that can not be mapped to an attribute is left as free token. In this paper, we assume access to a semantic parser such as  \cite{SPT10} to understand the semantics of the tokens, and focus on the problem of learning to reformulate queries that contain modifiers into ones that specify precise attribute values that can well satisfy the users.

When semantics is inferred for a subset of tokens, search is performed by combination of exact or approximate match of semantic tokens against the attribute values in the index  and keyword search against the textual description of the product (using the free tokens). As textual descriptions tend to be sparse or inaccurate, keyword search becomes ineffective.  Thus, the problem has attracted attention from both the IR as well as the database communities. There has been a number of studies on answering keyword queries for both traditional and XML databases~\cite{ACD02dbxplorer, ding+11top-k, guo+03xrank, HGP03ir-database, HP03keyword-database, KZG09a, KZG09,lin+08text-cube, liu+06effective, luo+07spark, XHG10}.  A recent survey is given in~\cite{YQC09keyword-database}.  They generally focus on three questions: how to efficiently retrieving tuples that contain the keywords from the database, how to find relations that can be joined to produce such tuples, and how to support efficient query processing that involves retrieval functions such as BM25.  An essential assumption is that the keywords being searched appear in the database somewhere. Our work addresses a different problem, where the keywords we are interested in may not be explicitly mentioned in the database, and a mapping from keywords to attribute-value pairs has to be learned through user queries and their browse trails.  Our work can complement, for example, the review-driven approach in~\cite{ding+11top-k} by providing an additional source of signals from user behavior.

Related work in web search is primarily on query reformulations~\cite{LWA09, ERW11} and document expansion~\cite{CGX09,ACCGKX09}. Our work is primarily different in the sense that we study the mapping of free tokens in the query to structured attribute values describing the products in the catalog.



\section{Model and Formulation} \label{sec:model}

\subsection{Overview}

Our work is done in the context of a commerce search engine that answers commerce queries given a product catalog.  We assume the existence of a semantic parser that identifies the attributes in the query and extracts their associated values, based on past work such as~\cite{SPT10}.  We call the processed queries \emph{annotated queries}.  For example, the query \query{gucci leather handbags} will be annotated as \query{category:handbags brand:gucci material:leather}.\\However, not all tokens in the query can be matched to an attribute.  In particular, users may use terms such as \texttt{designer} that do not match any attribute value in the product catalog.  Such tokens will be marked as \texttt{free} by the semantic parser.  The goal of our work is to understand how to identify the free tokens that change the sets of products to be retrieved, called \emph{modifiers}, and to learn to reformulate queries that contain modifiers into ones that specify precise attribute values that can well satisfy the users, such as from \query{category:handbags free:designer} to \query{category:handbags brand:gucci material:leather}.

At a high level, our solution is to analyze the user browse sessions to discover common features of the products that can satisfy the modifiers.  It consists of four steps.  First, we analyze the sessions to generate labels for domains.  Second, we identify the set of valid modifiers among all tokens that we cannot map to any product attributes.  Third, we estimate the likelihood of each modifier being associated with particular attribute values.  Finally, we retrieve products from a database based on the identified attribute values, and generate query rewrites that are good representations of the retrieved products.  We will go into each of these steps in further details in this section.  First, we introduce notations that will be used throughout the section.

Let $P$ denote a database of products.  Let $A$ denote the set of attributes of the products, with $|A| = k$.  For each attribute $a \in A$, let $V_a$ denote the set of valid values for attribute $a$.  We denote the set of all valid attribute-value pairs, or \emph{AV pairs} for short, by $AV$, defined by the set $\{(a, v) : a \in A, v \in V_a\}$.  Each product $p \in P$ is represented as a set of AV pairs, $\{(a_1, v_1), (a_2, v_2), \ldots, (a_k, v_k)\}$, one for each attribute.  For a set of AV pairs $S$, let $P(S)$ denote the subset of products in $P$ that match all of the specified attribute values.  For example, if $S = \{\av{brand, sony}, \av{diagonal size, 32}\}$, then $P(S)$ represents all products with brand equals sony and with diagonal size equals 32.

Let $U$ denote the collection of user browse sessions.  We call these sessions \emph{browse trails}, where each trail is associated with a query and a sequence of websites visited by the user.  As we will be analyzing these trails at the domain level, let $D$ denote the set of domains.  We denote the collection of all tokens in the queries by $T$.  These tokens include both \emph{typed tokens} expressed as AV pairs and \emph{free tokens} that the parser cannot map to any attribute value, denoted by $F$.  We consider some of these free tokens to be \emph{modifiers}, i.e., tokens that restrict the subset of products to be retrieved.  We denote the set of modifiers by $M$.  Stated formally, the goal of our work is to find an algorithm that associates each modifier $m \in M$ with a set of AV pairs $S \subset AV$.  Success will be measured empirically via human judgements, as described further in Section~\ref{sec:experiments}.

\subsection{Labeling Domains} \label{ssec:browsetrails}

In order to discover the product features a user is interested in given a query containing a modifier, we start by discovering the products the user examined after her query.  As search engines may fail to understand the query and do not surface the right results, we consider not only the page that a user clicked on but also the subsequent pages the user visited in the session.  Such collection of interactions have been used in previous studies and are called \emph{browse trails}~\cite{BW08,PG11,WD07}.  Using browse trails is especially important for understanding queries that contain modifiers, as our experience suggests that search engines are especially poor in answering such queries.

Ideally, we would like to find out the exact product(s) on each of the webpages on the browse trail; unfortunately, due to the large volume of data that needs to be processed, it is computationally infeasible to parse each of the webpages.  Instead, we label the websites by propagating the query tokens, including both AV pairs and free tokens, along the browse trails.  Further, to overcome data sparsity, we group websites by their host domains and generate labels at the domain level.  Intuitively, this is a reasonable approach because users looking for \query{sony tv} will likely spend more time on \url{http://www.sony.com} than, say, \url{http://www.frys.com}, whereas users looking for \query{52 inch lcd tv} will spend more time on general merchant sites such as \url{http://www.bestbuy.com} and \url{http://www.nextag.com}.  Likewise, users looking for \query{widescreen tv} will more often end up on merchants that have widescreen televisions in their catalog.  A similar approach has been previously proposed in~\cite{PG11} and has been shown to produce good labels for domains.

Stated formally, the goal of our first step is to take as input the collection of browse trails $U$ and output the frequency counts $c(t,d)$ of how often queries containing token $t \in T$ ends up visiting domain $d \in D$.  The frequency counts are computed using a variant of the heavy hitter algorithm.  In short, for each trail $u$, we create a set of token-domain pairs by pairing up each domain in $u$ with each of the tokens in the query that $u$ originated from.  We then ran a sampling-based algorithm to determine approximate counts.  Through careful accounting, the resulting frequency counts can be shown to be close to the true counts~\cite{PG11}.  Given these counts, together with suitable normalization, we can compute for example the distribution of free tokens for a given domain $d$, $\Prob(f | d)$ for $f \in F$.

We illustrate the output of this computation through an example from the televisions category.  In Table~\ref{tab:domain_avPairs}, we show the probability distribution of different brands and model numbers for three domains.  These values tell us that visitors to the domain \url{www.target.com} often start with a query that specify a manufacturer of smaller-sized TVs (Westinghouse and Haier produce many portable TVs), whereas visitors to \url{www.avsforum.com} are more mixed (Vizio and Pioneer produce TVs of all sizes). In addition, users visiting \url{www.target.com} typically do not begin their queries with model numbers, while those visiting \url{www.avsforums.com} do, and coincidentally with model numbers that correspond to televisions with large screen sizes.  

Likewise, we can investigate the modifiers associated with each domain.  In Table~\ref{tab:domain_portable}, we show the domains and their association with the modifier \texttt{portable}, as measured by $\Prob(\texttt{portable}|d)$, which can be interpreted as among all modifiers that are associated with a domain, what is the fraction of tokens that equal \texttt{portable}.  We see that the domain \url{www.target.com} has the largest support for this modifier, while the forum site \url{www.avsforum.com} has relatively little support.

\begin{table}
\small
\begin{center}
\begin{tabular}{llrlr}
\FL
Domain $d$ & Brand $b$ & $\Prob(b | d)$ & Model $n$ & $\Prob(n | d)$
\ML
\multirow{5}{*}{\url{www.target.com}}
  &Westinghouse& 0.38 & &\\
  &Vizio& 0.33 & &\\
  &Haier& 0.07 & &\\
  &Sylvania& 0.06 & &\\
  &Magnavox& 0.05 & &
\ML
\multirow{5}{*}{\url{www.sears.com}}
  &Vizio& 0.54 & tc-p54s1 & 0.24\\
  &Sylvania& 0.17 & tc-p42x1 & 0.16\\
  &Zenith& 0.10 & tc-p42s1 & 0.12\\
  &Coby& 0.06 & kdl52s5100 & 0.08\\
  &Haier& 0.04 & ln40b530 & 0.07
\ML
\multirow{5}{*}{\url{www.avsforum.com}}
  &Vizio& 0.51 & tc-p58v10 & 0.28\\
  &Pioneer& 0.22 & tc-p54s1 & 0.17\\
  &Panasonic& 0.15 & 55sv670u & 0.11\\
  &Dynex& 0.03 & pn58b650 & 0.10\\
  &Insignia& 0.02 & kdl52xbr9 & 0.07
\LL
\end{tabular}
\caption{Association between domains and AV pairs, grouped by attributes, in the television category} \label{tab:domain_avPairs}
\end{center}
\end{table}

\begin{table}
\begin{center}
\begin{tabular}{lr}
\FL
Domain $d$ & $\Prob(\texttt{portable} | d)$
\ML
\url{www.target.com} & 0.21 \\
\url{www.commentcamarche.net} & 0.09 \\
\url{www.sears.com} & 0.07 \\
\url{www.walmart.com} & 0.04 \\
\url{www.bizrate.com} & 0.02 \\
\url{www.avsforum.com} & 0.01
\LL
\end{tabular}
\caption{Association between domains and the free token \texttt{portable} in the television category}
\label{tab:domain_portable}
\end{center}
\end{table}

At this juncture, one may wonder if we can simply stop and select the domains that have high support for a free token, and select all the AV pairs that have high support in that domain to be the mapping of interest.  For example, in light of Tables~\ref{tab:domain_avPairs} and~\ref{tab:domain_portable}, for the free token \texttt{portable}, we include all the dominant AV pairs for \url{www.target.com}.  This approach has multiple issues.  First, choosing dominant domains, and subsequently the AV pairs associated with that domains requires the introduction of threshold parameters that require tuning.  Second, and more importantly, even if these threshold choices were made, the resulting mapping between the modifier and the AV pairs might lack generalization because the resulting mapping may restrict the choices (such as restricted to certain model numbers or brands).  For instance, in our running example, if only \url{www.target.com} was was selected, we will restrict the brands to Westinghouse and Haier, while brands such as Coby and Dynex also make portable televisions.

To work around this limitation, we instead treat the frequency counts computed in this step as input and estimate a conditional distribution of the AV pairs given the free tokens.  We then use the AV pairs with high conditional probabilities to retrieve products from the catalog. Finally, we use these products to identify the common features of the products and generate a reformulation for the modifier.

Before, we proceed to explain these steps, we will describe in the next section how we employ these frequency counts to identify the free tokens that correspond to modifiers, tokens that influence the set of products to be retrieved.

\subsection{Identifying Modifiers} \label{sec:vm}

The goal of this step is to identify the set of modifiers from amongst all free tokens.  Intuitively speaking, we consider a free token to be a modifier if it helps the user distinguish what kinds of products she has in mind.  As our analysis is aggregated at the domain level, we consider a free token to be distinguishing if the webpages the users went to are concentrated over few domains.

Drawing on this intuition, we propose a scoring mechanism based loosely on the TF-IDF retrieval function.  Specifically, for each free token $f \in F$, let $imp(f)$ denote its importance score, as given by
\[
  imp(f) = \sum_{d \in D} \Prob(f|d) \log\left(\frac{|D|}{1.0+df(f)}\right)
\]
where $df(f)$ is number of domains in $D$ for which the free token $f$ has a non-zero weight.  We then select the $10$ free tokens with the highest scores for each category as candidate modifiers.

Table~\ref{tab:vm} illustrates these modifiers ordered by their importance scores in a number of popular product categories.  As the table shows, our method identifies many terms that restrict the set of products to be retrieved at the top.  For example, we see modifiers like {\tt portable} (small and light in weight), {\tt streaming} (special feature) for televisions; and modifiers like {\tt evening} (restrictions on color and materials), {\tt small} (size restriction) for handbags etc.  A subset of these category-modifier pairs, together with modifiers for another six categories, will be used in our experiments in Section~\ref{sec:experiments}. In the next section, we present the details of computing the association between modifiers and AV pairs.

\begin{table*}
\begin{center}
\begin{tabular}{ll}
\FL
Category & Candidate modifiers
\ML
Refrigerators & commercial, compact, counter-depth, best, freezerless, outdoor, small, side-by-side, undercounter, efficient \\
Air Conditioners & central, ductless, home, remote, best, small, efficient, commercial, quiet, evaporative \\
Knives & electric, professional, gourmet, best, chef's, safe, home, ultimate, outdoor, expensive \\
Ovens & commercial, electric, single, standing, convection, freestanding, professional, outdoor, top, downdraft \\
Dishwashers & best, countertop, quiet, tall, efficient, compact, built-in, portable, home, professional \\
Handbags & evening, small, big, newest, popular, oversized, casual, exotic, designer, plus \\
Televisions & portable, remote, streaming, flat, largest, hd, biggest, compatible, thin, built-in \\
Radar Detectors & remote, newest, solar, portable, built-in, satellite, powerful, max, luxury, maximum \\
Voice Recorders & portable, tiny, interactive, remote, powerful, thin, cool, hd, deluxe, built-in \\
Jackets & kids, women's, girls, retro, hooded, insulated, running, maternity, designer, distressed
\LL
\end{tabular}
\caption{Top ten valid modifiers for different categories.} \label{tab:vm}
\end{center}
\end{table*}

\subsection{Estimating Association Probability} \label{ssec:covisitation}

The goal of this step is to estimate, for each modifier $m$,  the probability $s_i$ with which an AV pair $(a_i, v_i)$ is relevant to this modifier.  In order to compute this probability, we make use of the following observation. If a modifier is related to an attribute-value pair, it is very likely that queries that contain either of them will ultimately lead the users to the same domains. Therefore, we can leverage the browse trails to compute the association of an AV pair with a modifier, and normalize across domains to compute the required probabilities.

In particular, we postulate a generative process of modifiers and AV pairs as described in Algorithm~\ref{alg:gen}. A domain $d$ such as \url{www.sony.com} is chosen according to some prior probability over domains, $\Prob(d)$. Once the domain is chosen, the modifier such as \texttt{portable} becomes conditionally independent of the AV pairs. An AV pair $(a, v)$ is generated by first choosing an attribute according to the domain, and then choosing a value based jointly on that attribute and the domain. The values are conditioned on the attribute and the domain since value distribution is also influenced by the domain under consideration. For instance, for a brand-centric domain such as  \url{www.sony.com}, we would expect $\Prob(v|a,d)$ to peak at a particular value (in this example, \texttt{sony}) for attribute \texttt{manufacturer}, while for domains such as \url{www.walmart.com}, $\Prob(v|a,d)$ will be more uniform. Using this generative model, we can write the joint distribution as a product of conditional distributions defined by the generative process:
\begin{equation}
\Prob((a,v), m , d ) = \Prob(d)\Prob(a|d)\Prob(v|a,d)\Prob(m|d).  \label{eqn:joint}
\end{equation}
Each of these quantities can be directly obtained from the frequency counts computed in Section~\ref{ssec:browsetrails}.  

\begin{algorithm}[t]
Pick a domain $d$ according to $\Prob(d)$\\
Pick an attribute $a$ according to $\Prob(a|d)$ \\
Pick a value $v$ according to $\Prob(v|a,d)$ \\
Pick a modifier $m$ according to $\Prob(m|d)$ \\
\caption{Postulated generative model of modifiers and AV pairs.}  \label{alg:gen}
\end{algorithm}

By marginalizing the joint distribution with respect to the domains, we obtain
\begin{equation}
\Prob((a,v), m  ) = \sum_{d} \Prob(d)\Prob(a|d)\Prob(v|a,d)\Prob(m|d)
\end{equation}
The conditional probability of an AV pair $(a,v)$ given a modifier $m$, which we call the \emph{association score}, is given by the chain rule of probability, and equals:
\begin{equation}
\Prob((a,v)|m) = \frac{\Prob((a,v), m } { \sum_{m' \in M} \Prob((a,v) \in AV, m')}
\end{equation}

Continuing with our example, Table~\ref{tab:av_portable} shows a portion of the results at the end of this step for the modifier \texttt{portable}.  Note that the association scores approach has successfully identified brands that are the major manufacturers of portable televisions.  Later in the paper (Section~\ref{sec:numattr}), we will show that it can also successfully identify small diagonal sizes as being highly associated with \texttt{portable} after clustering diagonal sizes.  On the other hand, it has also identified certain model numbers that correspond to televisions with large screen sizes (\eg kdl40s5100) as being relevant for this modifier.

While this provides an initial candidate set of mappings between the modifier and the AV pairs, we would like to obtain those mappings that (a) bring in combinations of AV pairs that are more precise to the modifier, and (b) can generalize to the subspace of products that are relevant to the corresponding modifier. To do so, we make use of the product catalog as explained in the next step.

\begin{table}
\begin{center}
\begin{tabular}{lr}
\FL
AV pair $av$ & $\Prob(av | \texttt{portable})$
\ML
(brand, Audiovox) & 0.017 \\
(brand, Emerson) & 0.017 \\
(brand, Westinghouse) & 0.017 \\
(model, kdl40s5100) & 0.017 \\
(model, vf550m) & 0.017 \\
(brand, Haier) & 0.015 \\
(brand, Viore) & 0.014 \\
(brand, Sylvania) & 0.014
\LL
\end{tabular}
\caption{Conditional distribution of AV pairs given the modifier \texttt{portable} in the television category}
\label{tab:av_portable}
\end{center}
\end{table}

\subsection{Generating Rewrites} \label{ssec:itemset}

The final step of our algorithm produces sets of AV pairs that can satisfy the modifiers using the product catalog and the AV pairs with association scores as input.  This is an important step as the attribute values identified earlier in the process are limited to ones that appear in user queries; these terms are often skewed towards certain attributes depending on the category.  For example, in the category of electronic products, a large fraction of queries consist of solely a model number, and hence a modifier is often associated with a set of model numbers.  We can vastly improve recall by figuring out how these products manage to satisfy the query.

Intuitively speaking, we would like to find a set of AV pairs that is both specific, i.e., pinpoints as many attribute values as possible, especially for important attributes, and contains as large a fraction of the products that match the AV pairs with high association scores with the modifier as possible.  Further, observing that AV pairs that are common in the catalog are more likely to be associated with a modifier by random chance, we want to weight the association scores of the AV pairs inversely proportional to the number of the products with that AV pair in the catalog.  This leads us to the concept of \emph{coverage scores}.

\begin{definition}[Coverage Scores] \label{def:coverage-scores}
  Let the set of AV pairs with association scores be denoted by $C = \{(a_i, v_i, s_i)\}_{i=1}^m$.  We define the weight function $w : P \mapsto \Rea_+$ as
  \[
    w(p = \{(a_1, v_1)\}_{i=1}^k) = \sum_{(a, v, s) \in C : (a, v) \in p} \frac{s}{|P(\{(a, v)\})|}
  \]
  which measures how well product $p$ satisfies the set $C$.

  The {\bf coverage score} $s$ of a set of AV pairs $S = \{(a_i, v_i)\}_{i=1}^n$, with respect to a set of AV pairs with association scores $C$ and importance values of attributes $z : A \mapsto \Rea_+$, equals
  \[
    c(S) = w(S)z(S)
  \]
  where
  \begin{align*}
    w(S) &= \frac{\sum_{p \in P(S)} w(p)}{\sum_{p \in P} w(p)} \\
    z(S) &= \frac{\sum_{a : (a_i, \cdot) \in S} z(a)}{\sum_{a \in A} z(a)}
  \end{align*}
  for which $w(S)$ can be interpreted as the fraction of weights that the products satisfying $S$ covers among all products, and $z(S)$ the relative importance of the attributes covered in $S$.
\end{definition}

\begin{figure}
\begin{center}
\includegraphics[width=2.25in]{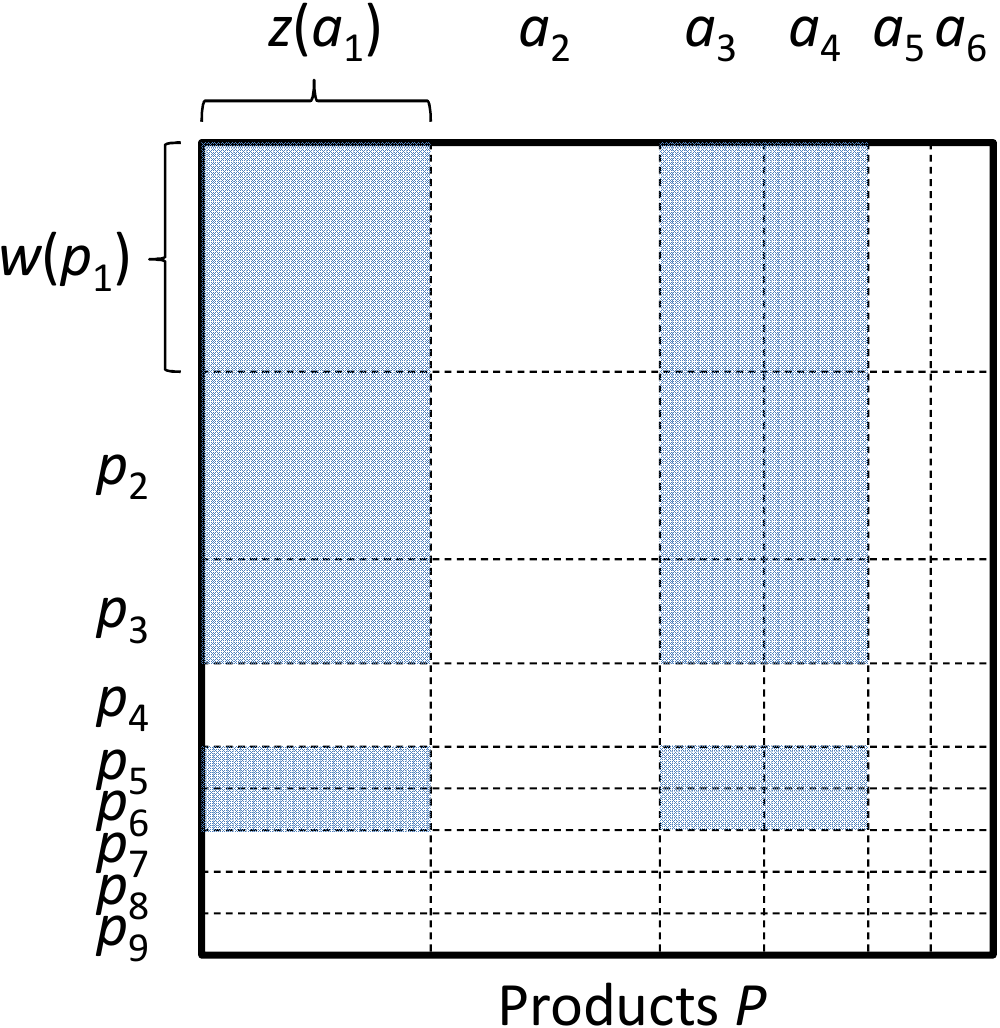}
\caption{The coverage score of a set of AV pairs that include attribute $a_1$, $a_3$, $a_4$, satisfied by products $p_1$, $p_2$, $p_3$, $p_5$, $p_6$}  \label{fig:coverage}
\end{center}
\end{figure}

We illustrate the concept of coverage scores in Figure~\ref{fig:coverage}.  In the figure, each row corresponds to a product in the database $P$.  A row is higher if it has more weights according to the weight function induced by the set $C$ of AV pairs and association scores computed in the last step.  Each column corresponds to an attribute.  A column is wider if the attribute is deemed more important by $z$.  The coverage score thus measures the fraction of area covered by a set of AV pairs in the space of all relevant products and attributes.

Unfortunately, we were unable to determine if there exists an efficient algorithm to find the set of AV pairs that maximizes the coverage score.  In this study, we develop a heuristic solution that draws on the ideas of finding frequent itemsets .  In our setting, an item corresponds to a particular AV pair, and an itemset corresponds to a set of AV pairs.  Typically, given a set of baskets  $B$ and a desired minimum support threshold $\theta$, an itemset algorithm finds all maximal itemsets with at least $\theta$ number of baskets that contain the itemset.  It is easy to adapt the apriori algorithm~\cite{AS94} for finding itemsets to incorporate weights.  Lemma~\ref{lem:coverage} connects the problems of finding frequent itemsets and of finding the set of AV pairs with the highest coverage scores.

\begin{lemma} \label{lem:coverage}
  A set of AV pairs with the highest coverage score must be a maximal itemset for some threshold $\theta$.
\end{lemma}

\begin{algorithm}[t]
\In{A set $C = \{(a_i, v_i, s_i)\}_{i=1}^m$ of AV pairs $(a_i, v_i)$ with association scores $s_i$, product database $P$, importance values of attributes $z$}
\Out{A set $\mathcal{S} = \{(S_j, c_j)\}_{j=1}^n$ of sets of AV pairs $S_j$ with coverage scores $c_j$}
\BlankLine
$P'' \leftarrow \emptyset, W \leftarrow 0$\;
\ForEach{$(a_i, v_i, s_i) \in C$}{
  $P' \leftarrow P(\{(a_i, v_i)\})$\;
  \ForEach{product $p \in P'$}{
    Add $p$ to database $P''$\;
    $w[p] \leftarrow w[p] + \frac{s_i}{|P'|}$\;
  }
  $W \leftarrow W + s_i$\;
}
$\mathcal{S} \leftarrow \emptyset$\;
\For{$\theta \in (0, 1)$}{
  $\mathcal{S}' \leftarrow$ \textsc{Find-Itemset}$(P'', w, \theta W)$\;
  \ForEach{$S \in \mathcal{S}'$}{
    $w(S) \leftarrow \sum_{p \in P(S)} w[p], z(S) \leftarrow \sum_{a : (a_i, \cdot) \in S} z(a)$\;
    $c \leftarrow w(S) \times z(S)$\;
    Add $(S, c)$ to $\mathcal{S}$\;
  }
}
\Return $\mathcal{S}$\;
\caption{Identify Product Features}  \label{alg:itemset}
\end{algorithm}

Putting it together, our algorithm for finding the set of AV pairs with the highest coverage score is given in Algorithm~\ref{alg:itemset}.  We start by retrieving the set of products $P'$ for each of the AV pairs identified in the previous step. For each retrieved product $p$, we insert it into a database $P''$, and increase its weight in the weight matrix $w$ by the association scores $s_i$ divided by the number of products satisfying the given AV pair.  We then loop through different values of support threshold ratios $\theta$, and invoke the adapted itemset algorithm that works on weighted database to find all maximal itemsets with minimum support ratios of $\theta$.  We compute the coverage scores for each of the itemsets, and add them to the output.  Since both denominators of $w(S)$ and $z(S)$ are constant relative to $C$, we only need to compute the numerator if we are interested in the relative ordering of the itemsets.  Note that the algorithm as stated solves the more general problem of \emph{computing} the coverage scores for sets of AV pairs.  We ended up solving this problem as we find it helpful in the experiments to combine these sets heuristically at retrieval time.

Based on Lemma~\ref{lem:coverage}, if we want to find the set of AV pairs with the highest coverage score, we will need to try all possible values of $\theta$, which is infeasible.  In the implementation of Algorithm~\ref{alg:itemset}, we conduct a $\epsilon$-grid search over the range $(0, 1)$, and hence cannot guarantee finding the absolute best set.  Nonetheless, in our experiments, we find that we end up with the same set regardless of the choice of the grid size $\epsilon$, provided $\epsilon$ is smaller than $0.1$.  An interesting research question is to establish a guarantee on the quality of the candidate set as a function of $\epsilon$.

\section{Experiments} \label{sec:experiments}

We conducted a series of experiments to evaluate our approach for rewriting queries that contain modifiers.  The first two experiments focus on the association scores, and evaluate the absolute and relative relevance of the selected attribute values.  The third experiment evaluates the end-to-end user experience of the search results retrieved based on the reformulated queries compared to that of using the original queries.

\subsection{Data Preparation}

We obtained the search and browse histories from consenting users of a popular browser toolbar over a 6-month period between November 2010 and April 2011.  We first classified the queries into product categories using a Na\"ive Bayes text classifier, and selected a number of categories of products for which there are an abundance of queries (over $10,000$ queries per category).  The queries from the selected categories were then parsed and annotated with type semantics on the techniques described in~\cite{SPT10}.  Next, we processed these histories as described in Section~\ref{ssec:browsetrails}. A set of valid modifiers was selected based on the criteria described in Section~\ref{sec:vm}.  The subset of modifiers and categories (from Table~\ref{tab:vm}) used in the experiments is given in Table~\ref{tab:cat-mod}.  We restricted the experiments only to these subsets based on the number of queries each modifier-category pair received and the uniqueness of the category under the top-level categories (of {\tt consumer electronics}, {\tt kitchen appliances}, and {\tt clothing and accessories}).

\begin{table}[t]
\begin{center}
\begin{tabular}{lll}
\FL
\textbf{Key} & \textbf{Category} & \textbf{Modifiers} \ML
\multicolumn{2}{l}{\textbf{Kitchen Appliances}} \\ \cmidrule{1-2}
AC & Airconditioners & central, commercial, ductless\\
DW & Dishwashers & portable, quiet\\
KN & Knives & chef's, gourmet\\
OV & Ovens & freestanding\\
RF & Refrigerators & counter-depth, small, undercounter
\ML
\multicolumn{2}{l}{\textbf{Clothing and Accessories}} \\ \cmidrule{1-2}
HB & Handbags & casual, designer, evening\\
JK & Jackets & designer, insulated, kids, retro\\
WL & Wallets & designer, stylish
\ML
\multicolumn{2}{l}{\textbf{Consumer Electronics}} \\ \cmidrule{1-2}
DP & DVD Players & portable, remote, streaming\\
RD & Radar Detectors & portable, remote\\
TV & Televisions & portable, remote, streaming\\
VR & Voice Recorders & portable, remote
\LL
\end{tabular}
\caption{Categories and modifiers evaluated in experiments.} \label{tab:cat-mod}
\end{center}
\end{table}

For each category, we retrieved product details from a product catalog where each product is specified as a set of attribute-value pairs as described in Section~\ref{sec:model}.  We then examined the distribution of attributes and kept only the attributes for which at least $> 10\%$ of the products have non-null values. Further, we restricted our analysis to only the categorical attributes such as brand, model, and color in a category. The reason behind this decision was due to the sparsity of numeric data in our catalog. We will show in Section~\ref{sec:numattr} how our approach can be extended to handle numeric attributes such as diagonal size and width. The resulting data after pre-processing constitutes the product database that we use throughout the rest of the section. 

\subsection{Evaluation of Identified Attribute Values}

In the first experiment, we evaluated the absolute relevance of the attribute values with high association scores for a modifier. The experiment is set up as follows.  For each modifier of each category, we selected the five values with the highest association scores for each attribute.  We then asked human judges to rate whether each of the selected attribute values are relevant to the modifier in question.  The judges are given three options: \texttt{relevant}, \texttt{not relevant}, and \texttt{unable to decide}.  Each attribute-value pair is evaluated seven times.  We post-process the results by filtering out any judgment that fails a simple sanity test.\footnote{We deliberately left some entries blank, and caught and threw out judges who assigned \texttt{relevant} to such entries.}

\begin{figure}
\begin{center}
\includegraphics[width=\columnwidth]{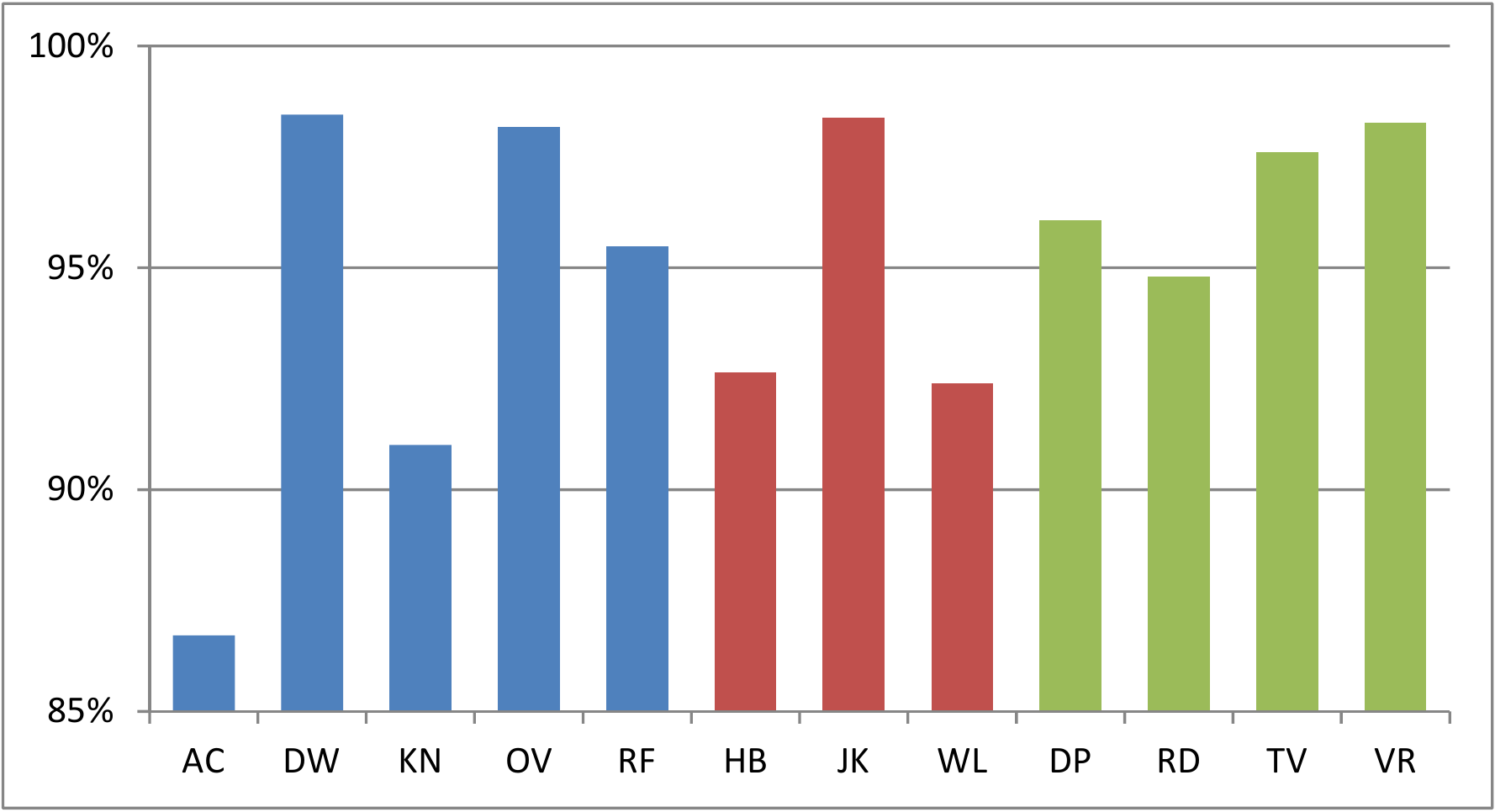}
\caption{Agreements rates of the selected attribute values by categories.  Categories that belong to the same major category are further grouped by colors.} \label{fig:agreement}
\end{center}
\end{figure}

We measured relevance by \emph{agreement rates}, defined as
\[
  \frac{\#\texttt{relevant}}{\#\texttt{relevant} + \#\texttt{not relevant}}
\]
The results by categories are shown in Figure~\ref{fig:agreement}.

Across the 12 categories, the average agreement rate of the selected attribute values is about $95\%$.  The results are consistent across the top-level categories, with the average agreement rate for kitchen appliances, clothing and accessories, and consumer electronics being $94\%$, $95\%$, and $97\%$ respectively.

We present in Table~\ref{tab:expt1} some anecdotal examples of the attribute values presented to human judges.  For {\tt portable dishwashers}, the association scores rank the top manufacturer to be {\tt Danby} followed by {\tt Edgestar}.  To verify this result, we manually examine a number of commerce portals and found that $24$" Danby dishwashers are considered a good candidate for portable dishwashers.  Likewise, we examine the manufacturers selected for evening handbags.  We found that Sydney Love handbags are typically more colorful and small compared to other manufacturers, two aspects that are usually associated with evening handbags.  In the category of refrigerators, we find that people who search for small refrigerators often look for wine chillers, beverage coolers etc.  Our results bear out our hypothesis. Finally, going through the products pages of dishwashers made by Dacor and made by Fisher and Paykel, we found words such as {\em WhisperWash} and {\em Quiet} that are used to describe the features of these dishwashers on Amazon.

\begin{table}
\small
\begin{center}
\begin{tabular}{lll}
\FL
\textbf{Modifier \& Category} & \textbf{Attribute} & \textbf{Attribute Values}
\ML
Ductless air conditioners & Manufacturer & Sanyo, Comfortaire \\
Portable dishwashers & Manufacturer & Danby, Edgestar \\
Quiet dishwashers & Manufacturer & Fisher and Paykel, GE, Dacor \\
Gourmet knives & Manufacturer & Lincoln, Stanley \\
Small refrigerators & Fridge Type & keg coolers, built-in \\
Small refrigerators & Manufacturer & Edgestar, Krups, Avanti\\
Designer handbags & Manufacturer & Chanel, Hermes, Balenciaga \\
Designer handbags & Bag Material & leather, fur \\
Evening handbags & Manufacturer & Sydney Love, Buxton \\
Evening handbags & Bag Material & straw, fabric \\
Designer hats & Product Name & Bermuda, Cowboy
\LL
\end{tabular}
\end{center}
\caption{Anecdotal examples of attribute values with high association scores for select modifiers}
\label{tab:expt1}
\end{table}

\subsection{Evaluation of Relative Ordering of Attribute Values}

In the second experiment, we evaluated whether attribute values with higher association scores are considered more relevant by the users than ones with lower scores.  The experiment is set up as follows.  For each modifier of each category, we selected the five attribute values with the highest and the lowest association scores.  We then created pairs of attribute values, one drawn from the top and another from the bottom, and asked human judges to rate which of the two values is more relevant.  Note that the bottom five attribute values are usually still relevant to the modifiers, as they have non-zero associations with the modifiers.  Therefore, an absolute test of whether an attribute value is relevant is inappropriate and insufficient, and a relative test is needed.

\begin{figure}
\begin{center}
\includegraphics[width=\columnwidth]{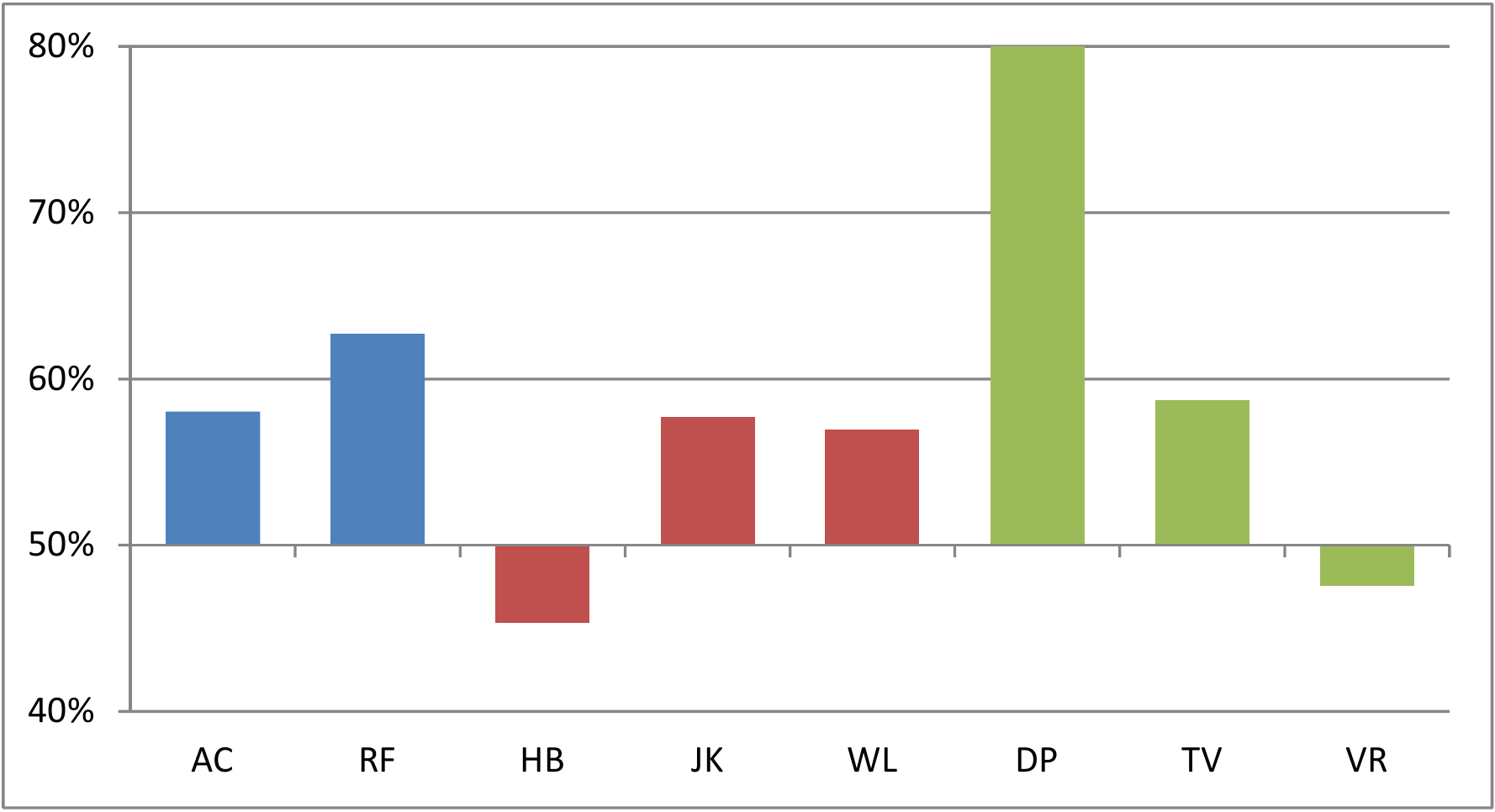}
\caption{Fraction of judgements that consider top five attribute values being more relevant than bottom five attribute values by categories. Categories that belong to the same major category are further grouped by colors.  Graph is centered at $50\%$, the expected result when the top five and the bottom five attribute values are equally relevant.} \label{fig:relative}
\end{center}
\end{figure}

We measured success by the fraction of the judges that find an attribute value from the top five is more relevant than one from the bottom five.  The results are shown in Figure~\ref{fig:relative}.  For ease of interpretation, we center the graph at $50\%$, the fraction one expects if there is no signal in the association scores.  Hence, a bar above the line indicates agreements with the scores, whereas below indicates disagreements with the scores.  Note that not all categories that appeared previously are present, as less than five attribute values were identified for some categories.

Across the eight categories, the average fraction of judges in favor of the top five attributes is $58\%$.  Given this fraction is computed with more than $2400$ observations, and the result is statistically significant under a one-proportion $z$-test ($p$-value $< 0.0001$) against the null hypothesis that the top and bottom five attribute values are equally good.  However, while we obtained good results for 6 out of 8 categories, we did poorly for handbags and voice recorders.  We examined the results in detail and found that one possible explanation is that the manufacturers with high association scores for {\tt evening handbags} and {\tt portable voice recorders} are typically lesser known brands that specialize in that specific segment of products, while the ones with low association scores are typically better known brands that produce all lines of products, and the judges consistently favor the better known brands.  We discuss this potential experimental bias further in Section~\ref{ssec:result-discussion}.

We present some of the anecdotal results that point to the difference between the attribute values with high and low association scores in Table~\ref{tab:expt2}.
In the top-level category of electronics, for dvd players, examining the product listings on Amazon suggests that {\tt Samsung} is more popular in selling streaming (i.e., wi-fi enabled) dvd players than {\tt Akai}; likewise, for televisions, {\tt Emerson} is better known for TVs with smaller diagonal size (20 inches and below) than {\tt LG electronics}, which produces televisions of all sizes, especially large LCD TVs.

\begin{table}[t]
\small
\begin{center}
\begin{tabular}{llll}
\FL
\textbf{Modifier \& Category} & \textbf{Attribute} & \textbf{High Assoc.} & \textbf{Low Assoc.}
\ML
Small refrigerators & Manufacturer & Edgestar & Princess \\
Undercounter refrigerators & Manufacturer & U-line & Zanussi\\
Evening	handbags & Manufacturer & Buxton & Hermes \\
Evening	handbags & Manufacturer & Sydney Love & Chanel \\
Stylish	jackets & Manufacturer & Joe rocket & Arc'teryx \\
Streaming dvdplayers & Manufacturer & Samsung & Akai \\
Portable televisions & Manufacturer & Emerson & LG
\LL
\end{tabular}
\end{center}
\caption{Anecdotal examples of attribute values with high and low association scores for select modifiers}
\label{tab:expt2}
\end{table}

\subsection{End-to-end Evaluation of Query Rewrites}

In the final experiment, we evaluated whether our query reformulation technique leads to improved relevance in the results retrieved.  The experiment is set up as follows.  For each modifier of each category, we consider the set of AV pairs with the highest coverage score as produced by Algorithm~\ref{alg:itemset}.\footnote{As part of the input to the computation of coverage scores, we need importance values for the attributes.  For this experiment, we treat all attributes as equally important.}  We generate a query reformulation by concatenating together the attribute values.  In many cases, we found that the coverage score is low due to a large number of missing values (nulls) in the database.  To address this database quality issue, we heuristically combine the sets of AV pairs with highest coverage scores for a modifier-category pair if the attributes are disjoint.  This heuristic is based on the observation that a missing value may take on any valid value from the domain, and hence it is sensible to combine multiple sets of AV pairs that are disjoint on the set of attributes selected.  The problem of data sparsity will be addressed further in Section~\ref{ssec:result-discussion}.  We issued both the reformulation and the original query to Amazon, and asked human judges to rate which of the results are more relevant to the original query.  The results are presented in Table~\ref{tab:end-to-end}.

\begin{table*}[t]
\begin{center}
\begin{tabular}{llll}
\FL
\textbf{Category} & \textbf{Original query} & \textbf{Rewritten query} & \textbf{Which query is better?}
\ML
AC & Ductless air conditioners & Sanyo mini split air conditioners & rewrite \\
AC & Commercial air conditioners & Haier portable air conditioners & rewrite\\
AC & Central air conditioners & Haier portable air conditioners & rewrite\\
DW & Quiet dishwashers & Maytag stainless steel dishwashers & original\\
DW & Portable dishwashers & General electric stainless steel dishwashers & rewrite\\
KN & Chefs knives & Wusthof knife sets & rewrite\\
KN & Gourmet knives & Wusthof knife sets & rewrite\\
OV & Freestanding ovens & General electric stainless steel ovens & original\\
RF & Small refrigerators & stainless steel Danby refrigerators & rewrite\\
RF & Counter-depth refrigerators & stainless steel Samsung refrigerators & rewrite
\ML
HB & Designer handbags & leather handbags & rewrite\\
HB & Evening handbags & Sydney Love fabric handbags & original\\
JK & Insulated jackets & waterproof jackets & rewrite\\
JK & Designer jackets & leather jackets & rewrite\\
JK & Kids jackets & waterproof jackets & rewrite
\ML
RD & Remote radar detectors & k-band city vg2 immunity radar detectors & rewrite\\
TV & Portable televisions & Samsung tft active matrix lcd hdtv & rewrite\\
VR & Portable voice recorders & Sony icd digital voice recorder & rewrite
\LL
\end{tabular}
\end{center}
\caption{End-to-end evaluation of query rewrites.} \label{tab:end-to-end}
\end{table*}

In $15$ out of $18$ queries ($87\%$), the judges prefer the reformulation over the original query.  Examining the result pages for both queries, we find that our results are better partly because of the products they retrieved, and partly due to the inclusion of solely products that belong to the category in question.  For example, the query \query{ductless air conditioners} retrieved a mix of air conditioners, books, and remote controls, whereas the reformulation \query{sanyo mini split air conditioners} retrieved only air conditioners.  This example highlights the danger of treating each query word as a keyword, as the keyword often cause unintended matches.

\subsection{Handling Numeric Attributes} \label{sec:numattr}

Thus far, we have only considered categorial attributes in our approach.  While in principle none of the four steps of our algorithm depends on attributes being categorical, in practice numeric attributes such as {\tt diagonal size} and {\tt price} pose additional challenges as they can take on many values.  A direct application of the algorithm that treat each different numeric value as unique will often result in extremely low association scores, and subsequently exclusion of these attributes from the reformulation due to low coverage scores.

To work around this problem, one should start by grouping the numeric values into a small number of buckets and treat all values within a bucket as equivalent.  The grouping can be done using standard database histogram techniques such as by equal width or equal depth.  The resulting buckets can then be treated as categorical attributes and our algorithm can proceed as before.

Due to data sparsity in our catalog, we did not manage to successfully apply this technique across all categories of products.  In some category, for example refrigerators, the availability of numeric data is strongly skewed towards the large refrigerators, rendering the technique inapplicable for modifiers that involve sizes.  Nonetheless, we found some success for some other categories.

For the category of TVs, we grouped the {\tt diagonal size} values into $5$ equal width buckets---0 to 40cm, 40 to 80cm, 80 to 120cm, 120 to 160cm, and 160cm and above; this width is selected to spread the TVs out well (approximately equal depth).  We then proceed with the algorithm and computed the association scores of diagonal sizes with the modifier \texttt{portable}.  The results are provided in Table~\ref{tab:ptv}.  As one can see, after grouping the diagonal sizes into $5$ sizes, our approach has selected the smaller TV diagonal sizes as being associated with \texttt{portable}.  We try varying the bucketing (for example to equal width buckets of 20cm instead of 40cm) and obtain similar results.

\begin{table}[t]
\begin{center}
\begin{tabular}{ll}
\FL
\textbf{Diagonal Size(cm)} & \textbf{Association score}
\ML
0 to 40 & 0.68 \\
40 to 80 & 0.65 \\
80 to 120 & 0.11
\LL
\end{tabular}
\end{center}
\caption{Mapping of {\tt portable} televisions to diagonal sizes} \label{tab:ptv}
\end{table}

For another category, we consider {\tt air conditioners}.  The numeric attributes of interest in this category are {\tt power output}. Consider {\tt central air conditioners}.  Table~\ref{tab:cac} shows the mapping of the modifier to the attribute values.  The results confirmed our intuition that users looking for {\tt central} air conditioners tend to look for ones with high power output than say compared to room air conditioners such as mini-split air conditioners.

\begin{table}[t]
\begin{center}
\begin{tabular}{llll}
\FL
\textbf{Power Output (BTU)} & \textbf{Association Score}
\ML
$\ge 15000$ & 0.22\\
12000 to 15000 & 0.16\\
8000  to 12000 & 0.15
\LL
\end{tabular}
\end{center}
\caption{Mapping of {\tt central} air conditioners to power output} \label{tab:cac}
\end{table}

\subsection{Discussion of Results} \label{ssec:result-discussion}

Our approach to inferring and associating AV pairs to modifiers is based on wisdom of crowds through the use of browse trails. Using the browse trails of a large number of users, we associated modifiers to the attribute value pairs, and used these associations along with their probabilities to infer sets of AV pairs that best describe the semantics.

Our first experiment showed that the top scoring attribute-value pair associations are highly relevant to the modifier. This validates our assumption that the domains that the users reach are likely to be similar for the queries that contain a modifier and its associated attribute-value pairs. Therefore, by tracing the trails of domains, one can find reliable associations. Further, our technique, for each modifier, is also accurate in determining the ordering of these associations, as shown by our second experiment. The third experiment shows the importance of generalizing the associations to enable better recall, both by adding additional attribute value pairs not present in the queries, and by identifying more holistic representation of the associations.  In 87\% of the cases, the re-written query using our approach resulted in retrieval of products considered more relevant by the users.

There are certain limitations to the experiments presented.  First, Mechanical Turk experiments are often noisy, and human judges could be subject to different sources of biases.  As noted before, we found that the human judges in the experiments have exhibited a bias towards better known brands.  This bias does \emph{not} necessarily work in our favor as our approach is not designed to take advantage of popular brands.  Indeed, for a number of cases our algorithm ends up selecting lesser-known specialty brands over well-known brands.  Second, human judges may not be always knowledgeable about the particular products.  As discussed in the Introduction, users issue queries containing modifiers partly due to lack of domain knowledge and do so to seek help from the search engine.  They may be unfamiliar with the manufacturers that specialize in making streaming DVD players or in making quiet dishwashers.  However, note that we are not asking the human judges to come up with the mapping, but rather to validate the algorithmically generated mapping, a relatively simpler task.  Further, we gave the option of \texttt{unable to decide} to the human judges, and our results are aggregated over many judgments.  As a further safeguard, we complement the Mechanical Turk experiment with a careful manual examination of a sample of the selected attribute values, and confirm that most of these attribute values are highly related to the given modifiers.

Finally, we would like to observe that our solution takes as input from several components and its ultimate success depends on the precision of these components.  Problems in these components can manifest itself in a variety of ways.  For example, queries can sometimes be misclassified, and annotations may confuse one attribute with another.  Such errors often lead to poor estimates of association scores.  The product catalog we work with has many missing values.  This limits the success of our final reformulation step as all the coverage scores were depressed.  Improving the quality of the catalog will certainly improve our results, and this was confirmed in a smaller-scale experiment where we manually scrubbed the air conditioner database and obtained better results.  Nonetheless, despite these limitations, our empirical experiment suggests that our approach can generate good query reformulations, and validates our belief that there are signals in the browse trails that can be harnessed to address the challenging problem of serving queries with modifiers.



\section{Conclusions}
\label{sec:conclusion}
We study the problem of query reformulation in commerce search that map queries containing modifiers to ones that specify precise attribute values of the products to be retrieved.  We did so by combining user behavior data and the product catalog of a commerce search engine to produce the mappings.  The user behavior data provides us with an initial association of attribute values with modifiers, the signal of which was then amplified and generalized through the use of the product catalog to identify common features of the products that satisfy the selected attribute values.  As part of a comprehensive user study, we find that users agree with the attribute values selected by our approach in about 95\% of the cases and they prefer the results surfaced for our reformulated queries to ones for the original queries in 87\% of the time.

There are several future research directions suggested by this study.  First, our work has focused on approaching the challenge of answering queries containing modifiers through query reformulation.  This was done in the context of treating the existing backend serving infrastructure of commerce search engines as given.  An interesting challenge is to develop an end-to-end solution that directly retrieves products for the given query.  We believe it is possible to develop a good solution by intelligently aggregating the top few sets of AV pairs with highest coverage scores.  Second, our solution does not explicitly take into account the noise introduced by the components it relies on nor the data sparsity of the catalog.  As explained in Section~\ref{ssec:result-discussion}, these have negative impacts on the quality of the rewrites.  While improving the data quality or the component quality is outside of the scope of this study, we believe it may be possible to produce better rewrites if these considerations are incorporated as part of a probabilistic framework.  Finally, there are other search domains where structured data exist and where it is common for users to issue queries containing terms that cannot be directly associated with the structure data, for example, in travel and in health.  It will be an important challenge to extend the techniques presented in this work to these other structured domains. 

\bibliographystyle{plain}
\bibliography{paper}

\end{document}